\documentclass[sigconf]{acmart}
\AtBeginDocument{%
  }

\setcopyright{acmlicensed}
\copyrightyear{2025}
\acmYear{2025}
\acmDOI{3715275.3732107}
\acmConference[FAccT '25]{}{June 23-26, 2025}{Athens, Greece}

\begin{document}

\title{Deepfakes on Demand: the rise of accessible non-consensual deepfake image generators}

\author{Will Hawkins}
\affiliation{%
  \institution{Oxford Internet Institute, University of Oxford}
  \city{Oxford}
  \country{United Kingdom}}
  \email{william.hawkins@wolfson.ox.ac.uk}
\orcid{0009-0004-5135-6792}

\author{Chris Russell}
\affiliation{%
  \institution{Oxford Internet Institute, University of Oxford}
  \city{Oxford}
  \country{United Kingdom}}
\email{chris.russell@oii.ox.ac.uk}

\author{Brent Mittelstadt}
\affiliation{%
  \institution{Oxford Internet Institute, University of Oxford}
  \city{Oxford}
  \country{United Kingdom}}
\email{brent.mittelstadt@oii.ox.ac.uk}


\begin{abstract}
Advances in multimodal machine learning have made text-to-image (T2I) models increasingly accessible and popular. However, T2I models introduce risks such as the generation of non-consensual depictions of identifiable individuals, otherwise known as \textit{deepfakes}. This paper presents an empirical study exploring the accessibility of deepfake model variants online. Through a metadata analysis of thousands of publicly downloadable model variants on two popular repositories, Hugging Face and Civitai, we demonstrate a huge rise in easily accessible deepfake models. Almost 35,000 examples of publicly downloadable deepfake model variants are identified, primarily hosted on Civitai. These deepfake models have been downloaded almost 15 million times since November 2022, with the models targeting a range of individuals from global celebrities to Instagram users with under 10,000 followers. Both Stable Diffusion and Flux models are used for the creation of deepfake models, with 96\% of these targeting women and many signalling intent to generate non-consensual intimate imagery (NCII). Deepfake model variants are often created via the parameter-efficient fine-tuning technique known as low rank adaptation (LoRA), requiring as few as 20 images, 24GB VRAM, and 15 minutes of time, making this process widely accessible via consumer-grade computers. Despite these models violating the Terms of Service of hosting platforms, and regulation seeking to prevent dissemination, these results emphasise the pressing need for greater action to be taken against the creation of deepfakes and NCII. 
\end{abstract}



\received{22 January 2025}
\received[revised]{N/A}
\received[accepted]{N/A}

\maketitle

\section{Introduction}
In the past five years, the development of generative text-to-image (T2I) models has evolved from a research domain to a global ecosystem of competitor products. Companies such as OpenAI, Google, Midjourney, Black Forest Labs, and Stability AI have led the industry in developing image generators which are increasingly indistinguishable from natural imagery. This has led to an increase in popularity across use cases, including advertising, animation, art, and design. In parallel access to T2I models has expanded thanks to an active open model development community. However, T2I models have also raised significant societal concerns, particularly regarding the potential for misuse. This paper investigates one area of misuse; the non-consensual generation of identifiable people, or \textit{`deepfakes'}.

The accessibility of T2I models has grown rapidly since the release of Stable Diffusion in 2022, a family of \textit{`open models'}  downloadable onto a local device \cite{solaiman_gradient_2023}. Open models give users more control, allowing them to fine-tune with new data and create custom variants after downloading. This has become increasingly popular thanks to breakthroughs in Diffusion Transformers (DiT) and accompanying parameter-efficient fine-tuning techniques such as Low Rank Adaptation (LoRA). These techniques enable a user to efficiently fine-tune a model whilst saving memory, improving the model on a specific task, whilst maintaining the overall quality of outputs. 

Greater model accessibility, despite its benefits, has also led to increased misuse, including a significant increase in deepfakes, particularly targeting sexually explicit content. Reports have found that the number of sexually explicit deepfake images has  risen rapidly year-on-year, increasing by 87 per cent  from 2022 to 2023 \cite{activefence_ai_2023}. This is increasingly impacting children, with 15 percent of high school students in the United States having heard about a sexual deepfake of someone associated with their school, and 13 percent of teenagers in the United Kingdom having experienced a sexually explicit deepfake image \cite{internet_matters_report_2024, wong_high_2024}. The issue overwhelmingly impacts women, with an estimated 99\% of sexual deepfakes targeting girls and women \cite{internet_matters_report_2024}.

Despite this increase and extensive coverage of the dissemination of AI-generated non-consensual intimate imagery (NCII), there is little literature studying the root cause of this issue: the underlying models and repositories that are used to produce this content. This paper bridges this gap by assessing how publicly downloadable T2I models may be used to create deepfake content. It explores to what extent two of the largest open model repositories, Civitai and Hugging Face, are used to host and share deepfake generators. It assesses two popular model families, Stable Diffusion and Flux, to understand whether popular T2I models are used for deepfake generation. Finally, we conduct metadata analysis to understand the extent to which model variant creators acknowledge, support, and/or inhibit the usage of their models to generate NCII.

We find that the availability of deepfake generators is increasing, identifying over 34,000 downloadable deepfake model variants intended to generate images of identifiable people. The overwhelming majority of these models are hosted on Civitai and have cumulatively been downloaded more than 15 million times since the launch of the platform in November 2022. Both Stable Diffusion and Flux are used to create deepfake model variants, with the release of Flux in August 2024 marking an acceleration in deepfake model creation. In a detailed analysis of 2,000 deepfake models, we find that 96\% target identifiable women, without any suggestion of consent being obtained from the subjects of these models. Metadata associated with models indicates that many are intended for the generation of sexual content, or non-consensual intimate imagery, despite this violating the Terms of Service of hosting platforms. We find that most deepfake model variants are created by training LoRA adapters, and can be developed with as few as 20 images, 24 GB of VRAM, and around 15 minutes of time, making this process widely accessible. These results are likely just the tip of the iceberg: each model downloaded could be used to generate potentially limitless deepfake images, and our analysis only considers publicly downloadable models rather than those fine-tuned and distributed within private communities. The findings suggest that the creation of deepfakes is more accessible and prevalent than ever before, and that model creators, developers, and regulators could do more to curb their growth and accessibility.  

\section{Related Work}
\subsection{Text-to-Image Models}

T2I generation has rapidly advanced within the last decade, thanks to breakthroughs in techniques including generative adversarial networks (GANs), transformer-based image generation, and diffusion models \cite{goodfellow_generative_2014, parmar_image_2018, sohl-dickstein_deep_2015}. These systems demonstrated the ability of a model to produce legible images from small amounts of text, with the diffusion-based approach leading a revolution in image generation capabilities. 

These research endeavours resulted in easily deployable systems with the release of DALL-E in 2021, an autoregressive transformer model developed by OpenAI \cite{ramesh_zero-shot_2021}. A year later, OpenAI followed this up with the release of DALL-E 2, where the use of diffusion models led to substantial quality improvements \cite{ramesh_hierarchical_2022}. Diffusion models add noise to an image and then learn to reverse this process, denoising the visual based on a textual prompt, thus creating an image for the user \cite{ho_denoising_2020}. Following DALL-E 2, a range of competitor models were released including Midjourney, Imagen 2, and Stable Diffusion which all utilised the diffusion architecture
\cite{imagen-team-google_imagen_2024, midjourney_team_midjourney_2025, rombach_high-resolution_2022, saharia_photorealistic_2022}. This led to a race to create high quality T2I outputs, increasing competition to create synthetic imagery which are almost indistinguishable from `real' images.

These advances in quality and ease of use have led to a variety of applications across creative domains. Initial models such as DALL-E and Midjourney were popular among artistic communities as a way to create and share AI-generated artworks \cite{ko_large-scale_2023}. They have since proven to be useful for a variety of tasks, including urban architecture, animation, education, and e-commerce advertising \cite{jiang_text2human_2022, seneviratne_dalle-urban_2022, vartiainen_using_2023, wang_disentangled_2025, zhao_enhancing_2024}. The versatility and rapid advancement of T2I models have led to economic opportunities for developers who create the most powerful and flexible models, increasing demand for high-quality models. 

\subsection{Open Models}
Today, access to leading T2I models ranges between those considered \textit{`closed'} and those which are \textit{`open'} \cite{solaiman_gradient_2023}. Closed models are generally only accessible via mechanisms mediated by the model developer, for example via an API or web-based user interface, where all engagement with the model is subject to control by the model creator. In contrast, open models are characterised by model weights being downloadable, meaning a user can run and modify the model on a local device without the oversight of the model developer. 

Stable Diffusion is perhaps the most prominent example of a family of open models, with various iterations of Stable Diffusion models being released by Stability AI since 2021. Stable Diffusion models can be accessed via model hosting platforms such as Hugging Face or Civitai, and downloaded to a local device. At the time of writing, the two most popular Stable Diffusion variants on Hugging Face, stable-diffusion-v1-5 and stable-diffusion-xl-base-1.0, had 5 million monthly combined downloads underlining their popularity \cite{huggingface_model_2024}. In August 2024 Black Forest Labs released a new T2I open model built using the DiT architecture, Flux, which has rapidly become popular with the open model community \cite{black_forest_labs_announcing_2024}. At the time of writing, the most popular variant of the Flux model hosted on Hugging Face had been downloaded over 1.2 million times in the last month, with over 3,000 variants of Flux models available to download on the platform \cite{huggingface_model_2024}. 

Open models have been popular among development communities due to the ability to use and fine-tune models locally \cite{han_svdiff_2023, ruiz_dreambooth_2023, xie_difffit_2023}. Fine-tuning involves taking a base model and conducting further training based on a defined dataset within a target domain. For example, if a user aimed for a model to become more capable at generating images of livestock they might choose to fine-tune on an image dataset of cows. Fine-tuning diffusion models has historically been computationally expensive and difficult to control, leading to few open models being widely fine-tuned prior to 2024 \cite{han_svdiff_2023}. 

The development of transformer-based architectures such as Diffusion Transformers and Vision Transformers has led to further advancements in the quality and customisability of open models \cite{dehghani_scaling_2023, peebles_scalable_2023}. Recent Flux and Stable Diffusion models have been built utilising transformer architectures, enabling fine-tuning via techniques such as LoRA \cite{chen_pixart-_2023, hu_lora_2021, huang_-context_2024, black_forest_labs_announcing_2024, esser_scaling_2024}. LoRA allows users to adapt pre-trained models to specific styles or concepts by introducing small, trainable adapter matrices into the model's layers. During fine-tuning, only these low-rank matrices are updated, while the original model weights remain frozen \cite{hu_lora_2021}. This process enables the model to learn new information from a specialised dataset without modifying the bulk of its parameters, significantly reducing the memory and computational resources needed compared to full fine-tuning. As of January 2025 there are almost 1,000 unique Flux variants on the Hugging Face Model Hub tagged with the term `LoRA', indicating the rising prominence of LoRA fine-tuning \cite{huggingface_model_2024}. 

\subsection{Text-to-image risks}
Whilst T2I models have demonstrated a wide range of uses and open T2I models are gaining popularity, they can also cause material societal harm \cite{bariach2024harmstaxonomyailikeness}. T2I models are trained on vast amounts of training data usually derived from the internet, which can lead to models perpetuating or amplifying stereotypes or biases found in the data \cite{birhane2023laionsdeninvestigatinghate, birhane_multimodal_2021, naik_social_2023, paullada_data_2021}. These issues can lead to harm amplification, where an output image is more harmful than an input prompt \cite{hao_safety_2023, hao_harm_2024}. Hao et al. demonstrate this issue in the domain of oversexualisation, showing that images of perceived females are sexualised at a higher rate than perceived males \cite{hao_harm_2024}. This may be due to training datasets containing images of a sexual nature and demonstrates that models are capable of generating sexually explicit imagery. 

T2I models can also be susceptible to adversarial users seeking to violate privacy norms. By fine-tuning using depictions of identifiable individuals, users can create deepfakes: synthetically generated content of identifiable people that appears entirely authentic \cite{gambin_deepfakes_2024, mirsky_creation_2022}. Models dedicated to deepfake generation can be created without an individual’s consent. Once a deepfake model of an individual is created there is potentially a limitless number of images which can be created. This has led to concerns relating to misinformation, efforts to prevent generation of identifiable individuals, and  watermarking of AI-generated content to track synthetic media \cite{dathathri_scalable_2024, hwang_brief_2023}. However, it is difficult for open model developers to completely prevent the creation of deepfakes due to the effectiveness of fine-tuning. 

In combination, the potential to generate sexual content and the possibility of creating images of identifiable individuals allows for the generation of NCII \cite{viola_designed_2023}. The demand for NCII is not new, with tools such as Photoshop previously used to create this content \cite{velez_why_2019}. There are online communities dedicated to the dissemination of deepfake content, with one platform hosting threads explicitly targeting `celebrity AI' representations \cite{4chan_4chan_2025}. AI-generated NCII has risen dramatically in popularity in recent years, with one popular deepfake pornography website receiving up to 17 million hits per month, with content almost exclusively targeting women \cite{tenbarge_found_2023}. These non-consensual depictions, sexual or otherwise, are often more difficult to create with closed models due to output filters \cite{imagen-team-google_imagen_2024, openai_team_dalle_2023}. Filter protections are often ineffective in open model domains, because users can choose to deploy a model on a local device without restrictions.

Regulatory proposals have come in response to the growth of deepfakes and NCII. The European Union (EU) AI Act places requirements on providers of AI systems capable of generating synthetic media to use provenance techniques such as watermarking or metadata identifiers to ensure AI-generated content can be consistently identified. Providers of AI agents must also disclose to users that they are interacting with an AI system \cite{european_union_regulation_2024}. The EU’s Digital Services Act (DSA) requires providers who moderate user-generated content to enforce notice-and-takedown procedures \cite{EU_DSA_2022}. In the UK (United Kingdom) the Online Safety Act has imposed a prohibition on deepfake dissemination, which has prompted criticism for regulating only the dissemination, but not the generation, of such content \cite{kira_when_2024, mcglynn_deepfake_2024, uk_government_online_2023}. More recently, the UK government has signalled intention to make creating sexually explicit ‘deepfake’ images a criminal offence \cite{uk_ministry_of_justice_government_2025}. A draft United Nations Cybercrime Convention proposed to ban the sharing of NCII, however this does not refer directly to AI-generated content, or propose to tackle its creation \cite{united_nations_draft_2024}. In the United States various states have implemented regulations relating to AI-generated NCII, however no federal-level prohibition currently exists to regulate the generation of such content \cite{graham_deepfakes_2024}. As a result, the regulatory landscape remains nascent, with limited attempts to restrict the creation of models capable of generating deepfakes. This study measures the breadth and severity of this key gap in AI regulation. 

\section{Method}
We conduct a three part study to understand the extent to which model repositories and specific open T2I models contribute to the proliferation of deepfake content and AI-generated NCII.

\subsection{Part A: Platform Analysis - Civitai}
Part A of our study explores the extent to which deepfake model variants are hosted and shared on one model repository, or hosting platform, Civitai. We select Civitai to make this assessment due to its popularity with text-to-image model developers and users, with over 100,000 fine-tuned variants of Stable Diffusion identified on the platform in one study \cite{luo_stylus_2024}. Unlike other platforms, such as Hugging Face, Civitai additionally enables model creators to tag model variants with a wide variety of searchable metadata such as tags including ‘Celebrity’ enabling a comprehensive review of models hosted on the platform. 

Civitai’s ‘Celebrity’ tag indicates that a model variant is intended to be used to produce depictions of identifiable individuals. From an initial evaluation of this tag we find that these model variants almost exclusively seek to produce photorealistic imagery of identifiable individuals, and therefore models with this tag are considered to be deepfake model variants. Model variants within this tag are usually named explicitly after an identifiable individual. For example, a model variant fine-tuned to generate images of Taylor Swift might be named “Taylor Swift LORA”. Using the Civitai API we curate a dataset of all model variants tagged with ‘Celebrity’. 

To understand the purpose and popularity of deepfake variants on Civitai we subsequently analyse various metadata attached to each variant. To understand the extent to which deepfake model variants are intended for the generation of NCII we develop a list of potential metadata red flags which indicate a variant might be intended for sexual use. Red flags include tags such as “sexy” or “porn” or reference to “sex” or “pornography” in model descriptions. We rely on red flags contained in metadata to determine whether a model is intended to generate NCII because sexually explicit imagery of identifiable individuals is not permitted on the Civitai platform. The full list of red flags identified can be found in Appendix \ref{appendix:d}. We additionally consider usage statistics and upload dates to determine how popular deepfake generators are, and understand any temporal trends on the platform. 

\subsection{Part B: Model Analysis - Stable Diffusion \& Flux}
Having established how frequently Civitai is used to host and disseminate deepfake model variants we next seek to understand whether popular models are fine-tuned to create deepfakes. We do this via a more focused assessment of a smaller dataset of models to allow for more fine-grained examination. At the time of writing, two of the most popular T2I models are variants of Stable Diffusion and Flux. On the Hugging Face platform Stable Diffusion 1.5 base model has over 3 million monthly downloads, with Flux.1-Dev garnering 1.1 million monthly downloads \cite{huggingface_model_2024}. Flux is a relatively new model with one major accessible iteration (Flux.1), while Stable Diffusion is a family of models of different sizes and varieties. As a result of this popularity, we choose Flux and Stable Diffusion model families for examination. 

Model analysis is conducted across two platforms: Civitai and Hugging Face. Like Civitai, Hugging Face is a prominent model repository, with monthly model uploads increasing exponentially in recent years \cite{castano_exploring_2023, castano_analyzing_2024}. Hugging Face’s Model Hub houses over 1.2 million models trained for a range of tasks, and over 50,000 models dedicated to text-to-imagery. We choose to conduct the study across two platforms to understand if policies and practices differ across major hosting sites.

We create a dataset of model variants for detailed analysis by accessing the respective APIs of each platform and conducting a search of all text-to-image models based on model names. Variants are added to the dataset if the terms ‘Stable Diffusion’, ‘SD’, or ‘Flux’ are present within the model name. We then filter out models which are scarcely used (e.g. with fewer than 10 monthly downloads on Hugging Face, or 250 lifetime downloads on Civitai), and models which are replications or quantisations of base models. Additionally, Civitai analysis focuses solely on LoRA variants of Flux and Stable Diffusion models, following findings in Part A suggesting the majority of deepfake models are created using LoRA. The full filtering criteria can be found at Appendix \ref{appendix:a}. In total, 15,349 models are identified as in-scope, as seen in Table \ref{tab:scope}. 

\begin{table}
  \centering
  \begin{tabular}{crrr}
    \toprule
    \textbf{Platform} & \textbf{Flux} & \textbf{Stable Diffusion} & \textbf{Total}\\
    \midrule
    \textbf{Hugging Face} & 1,286 & 3,872 & 5,158\\
    \textbf{Civitai} & 3,072 & 7,119 & 10,191\\
    \textbf{Total} & 4,358 & 10,991 & 15,349\\
  \bottomrule
\end{tabular}
  \caption{Count of models labelled as Flux and Stable Diffusion in-scope of study across Hugging Face and Civitai.}
  \label{tab:scope}
\end{table}

We manually label each in-scope model variant to determine if the variant is intended to generate deepfakes via judgements of the model names and descriptions. This might be explicit via the name of an individual being explicitly stated (e.g. `Bryan Cranston'), or implicit, for example if the individual is referred to by the character they play (e.g. `Walter White'). For each entry identified as a deepfake variant we additionally label the perceived gender of the subject, based on the name.

We additionally label each in-scope model variant to identify whether it is explicitly intended to generate sexually explicit content based on the model name. This review measures how frequently model variants are openly advertised as being intended for sexually explicit imagery. Models labelled as intended for sexually explicit content need not necessarily generate deepfakes, as individuals generated may be entirely AI-generated. The sexually explicit category and the deepfake category are not mutually exclusive, models can be both advertised as being intended to generate deepfakes and sexually explicit content.  

The full labelling criteria, developed based on an initial pilot, can be found in Appendix \ref{appendix:c}. Throughout the study we do not generate any images directly using model variants due to ethics concerns associated with utilising deepfake models. 

\subsection{Part C: Terms of Service and Model Accessibility}
We finally consider the Terms of Service of these platforms, and seek to understand how easily deepfake models can be produced and hosted. This provides insight into whether deepfake model variants are intended to be hosted on these platforms and helps us understand the ease to which they can be developed and deployed.  

\section{Results}
\subsection{Part A Results: Platform Analysis - Civitai}
To assess the prevalence of deepfake models on public model repositories the Civitai platform is searched for models which are labelled as intended to generate a `Celebrity'. These tags, assigned by model creators, indicate that a model is intended to generate an identifiable individual. Table \ref{tab:civitdata} provides an overview of these models and associated metadata. 

\begin{table}[h]
    \centering
    \begin{tabular}{ll}
        \toprule
        \textbf{Metadata} & \textbf{Count} \\
        \midrule
        Count of models tagged as `Celebrity' & 34,439 \\
        Cumulative downloads of Celeb models & 14,908,183 \\
        Count of Celeb models tagged as `LoRA' & 27,549 (80.0\%) \\
        Count of Celeb models tagged as `sexy' & 4,998 (14.5\%) \\
        Count of Celeb models tagged as `Instagram' or `Insta' & 1,165 (3.4\%) \\
        Count of Celeb models tagged as `nude' & 77 (0.2\%) \\
        Count of Celeb models tagged as `nsfw' & 46 (0.1\%) \\
        Count of Celeb models tagged as `female' & 6,883 (20.0\%) \\
        Count of Celeb models tagged as `male' & 99 (0.3\%) \\
    \bottomrule
\end{tabular}
\caption{Metadata analysis of Civitai models tagged as seeking to generate a `Celebrity'}
\label{tab:civitdata}
\end{table}

As seen in Table \ref{tab:civitdata} deepfake models are extremely popular on the Civitai platform, with over 34,000 unique models seeking the generation of a `Celebrity', spanning almost 15 million downloads. Of these models, the vast majority (80\%) are fine-tuned variants utilising low rank adaptation (LoRA). A significant number appear to be fine-tuned to generate intimate imagery, with the term `sexy' appearing in 14.5\% of the tags for `Celebrity' model variants. Some models are also tagged with sexually explicit indicators such as `nsfw' (46 instances) and `nude' (0.13\%). 99 of these 'Celebrity' models are tagged with the term `Male' whilst 6,883 are labelled with `Female', suggesting that the vast majority of models target women.

Additionally, a material proportion of the Celebrity tagged models do not appear to relate to individuals who might be traditionally labelled as public figures (e.g., politicians, film actors, popular musicians). Over 1,000 "Celebrity" models are tagged with terms suggesting they are known for their presence on Instagram, 287 tagged with `TikTok', 317 `Twitch', and 504 `YouTube' or `YouTuber'. Through a qualitative search of these model variants they do not all appear to depict individuals who could be considered celebrities. For example, one individual targeted by a deepfake model and labelled with both `Instagram' and `Celebrity' had fewer than 10,000 followers on the Instagram platform. 

Many of these deepfake model variants are developed by a subset of users. 2,804 unique users have uploaded a model labelled as `Celebrity' on Civitai, meaning each user has uploaded an average of 12 models. However, there appears to be a core group of highly active deepfake developers on these platforms. One user alone is responsible for 975 of all deepfake model variant uploads.  There is also a long-tail of `low activity' developers: 1,397, or almost 50\%, of Civitai users who have uploaded a deepfake model have shared just a single variant. Developers on Civitai can also be paid for their creations, via a method known as `Buzz'. 5,000 Buzz can be purchased for \$5, with deepfake models cumulatively receiving almost 1.5 million Buzz in tips (roughly \$1,500 USD). 

\subsubsection{Analysis of popular models}
We analyse the top one hundred most downloaded models on Civitai tagged with the term `celebrity' to determine which identifiable individuals are most frequently targeted by deepfake models. Each deepfake subject identified is searched via Google to assess perceived gender and primary occupation, with results outlined in Tables \ref{tab:Civitai-gender} and \ref{tab:occupation}. 

\begin{table}[htbp]
    \centering
    \begin{tabular}{lr}
        \toprule
        \textbf{Perceived Gender} & \textbf{Count} \\
        \midrule
        Female           & 97    \\
        Male       & 1    \\
        N/A            & 2     \\
        \bottomrule
    \end{tabular}
    \caption{Perceived gender analysis of the subjects of the top 100 most downloaded Celebrity / deepfake models on Civitai.}
    \label{tab:Civitai-gender}
\end{table}

\begin{table}[htbp]
    \centering
    \begin{tabular}{lrr}
        \toprule
        \textbf{Occupation} & \textbf{Count} & \textbf{Download Count} \\
        \midrule
        TV / Film           & 46     & 585,740 \\
        Musician       & 25     & 427,616  \\
        Adult sexual content            & 14      & 184,266  \\
        Internet Personality & 13      & 144,416  \\
        Other             & 2      & 90,355   \\
        \bottomrule
    \end{tabular}
    \caption{Occupation analysis of the subjects of the top 100 most downloaded Celebrity / deepfake models on Civitai.}
    \label{tab:occupation}
\end{table}

97 of the 100 models analysed targeted identifiable women, with 1 targeting an identifiable man and the remaining two models targeting groups of people comprised of a mix of genders. This indicates women are much more likely to be the subject of deepfakes, a hypothesis further explored in the analysis conducted in Part B. 

Of the 100, 46 target women who work in the TV and Film industry, including Emma Watson, Scarlett Johansson, and Liu Yifei. Of the remainder, 25 target musicians, and an additional 14 target actors in adult sexual content. 13 models target internet personalities, for example influencers on platforms such as TikTok, Twitch, or Instagram. Finally, two models pertained to mixed groups of people without a single discernible occupation, for example, a model labelled `Famous People'. 

\subsubsection{Temporal analysis}
\begin{figure*}[htbp]
    \centering
    \includegraphics[width=0.8\textwidth]{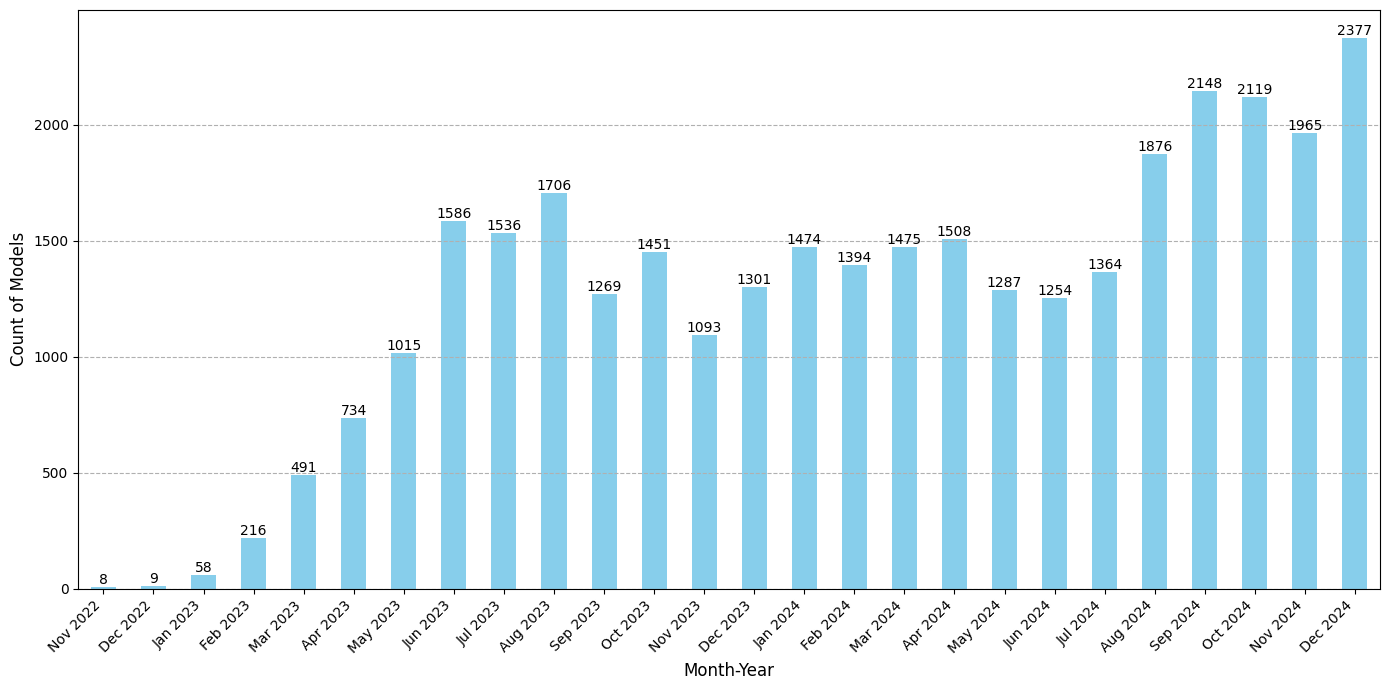}
    \caption{Count of models tagged ` Celebrity' per month on Civitai (up to December 2024).}
    \label{fig:temporal}
\end{figure*}

Temporal data provides further context on the extent to which the easy access to deepfake generators is a new phenomenon. As outlined in Figure \ref{fig:temporal}, the first model tagged with the `Celebrity' label appeared on Civitai in November 2022, the same month that the platform was launched. Since then, the numbers have ballooned, with December 2024 (the latest month captured) seeing the upload of almost 2,400 new deepfake model variants. Notably, August 2024 also appears to mark a substantial jump in the number of `Celebrity' deepfake models being posted to the hub. This coincides with the release of Flux, indicating that this model may have ushered in a new era of capabilities for individuals seeking to generate deepfakes of identifiable individuals. 

\subsection{Part B Results: Model Analysis - Stable Diffusion \& Flux}
These results establish that Civitai is used to share deepfake model variants, many of which appear intended to generate NCII. To further understand the types of models used to generate deepfakes we conduct a more detailed assessment of two model families, Flux and Stable Diffusion, spanning both the Hugging Face and Civitai platforms.

\subsubsection{Stable Diffusion}
Table \ref{tab:sd_count} reports deepfake and sexually explicit use cases for all variants labelled as Stable Diffusion models across both Hugging Face and Civitai. `Deepfake Models' are defined as those which explicitly target identifiable individuals. `Sexual Models' are model variants where the model title indicates that the use case is to generate sexual content, which may or may not be of an identifiable individual. The results show that while Stable Diffusion variants designed explicitly to create sexually explicit content do exist, they are only a small proportion (3.7\%) of all Stable Diffusion model variants uploaded to Hugging Face and Civitai. Stable Diffusion variants targeting only deepfakes form a larger, but still small proportion of the total models assessed (7.3\%). Civitai is found to host far more deepfake model variants, with almost 800 available on the platform, compared to just 11 on Hugging Face. 

We find a small number of Stable Diffusion variants (127) meet both the deepfake and sexually explicit criteria, usually due to references to individuals featuring in adult content. This is despite Part A of this study suggesting that many deepfake models on Civitai are intended for this use. These results may indicate that users developing deepfake models intended for sexual content generation may obfuscate the intended use cases to adhere to platform Terms of Service, explored further in Part C.

\begin{table}[h]
\centering
\begin{tabular}{lrrr}
\toprule
\textbf{Count of Models} & \textbf{Hugging Face} & \textbf{Civitai} & \textbf{Total} \\
\midrule
\textbf{Sexual Models} & 178 (4.6\%) & 233 (3.3\%) & 411 (3.7\%) \\
\textbf{Deepfake Models} & 11 (0.28\%) & 799 (11\%) & 810 (7.3\%) \\ 
\textbf{Total Models} & 3,872 (100\%) & 7,119 (100\%) & 10,991 (100\%) \\ 
\bottomrule
\end{tabular}
\caption{Count of Stable Diffusion models labelled as `sexual' or `deepfake' across platforms.}
\label{tab:sd_count}
\end{table}

Table \ref{tab:sd_usage} investigates the popularity of Stable Diffusion variants designed for sexual or deepfake content generation by measuring their usage statistics through download figures (which are measured as monthly downloads on Hugging Face and lifetime downloads on Civitai), as well as number of `likes' or `upvotes'. Sexual Models are a small but material portion of the total activity for Stable Diffusion models. On Hugging Face almost 10\% of analysed Stable Diffusion variant downloads pertain to Sexual Models with over 400,000 downloads per month on the platform. On the other hand, Deepfake Models are far more popular on Civitai than Hugging Face.

\begin{table*}[htbp]
\centering
\begin{tabular}{llrrr}
\toprule
\textbf{Platform} & \textbf{Usage Metric} & \textbf{Sexual} & \textbf{Deepfake} & \textbf{Total Usage} \\
\midrule
\textbf{Hugging Face} & Monthly Downloads & 412,207 (9.3\%)  & 2,280 (0.05\%) & 4,423,200 (100\%) \\
& Likes             & 552 (2.9\%)   & 20 (0.11\%)    & 18,897 (100\%)  \\
\midrule
\textbf{Civitai} & Total Downloads       & 305,355 (3.4\%)  & 759,049 (8.5\%) & 8,976,676 (100\%) \\
& Upvotes          & 33,986 (3.2\%) & 54,408 (5.1\%) & 1,073,952 (100\%) \\
\bottomrule
\end{tabular}
\caption{Stable Diffusion Sexual and Deepfake model downloads and likes/upvotes compared with total on Hugging Face \& Civitai.}
\label{tab:sd_usage}
\end{table*}

\subsubsection{Flux}
Flux is a comparatively newer family of models compared with Stable Diffusion, having initially been launched in August 2024.

\begin{table*}[htbp]
\centering
\begin{tabular}{lrrr}
\toprule
\textbf{Count of Models} & \textbf{Hugging Face} & \textbf{Civitai} & \textbf{Total} \\
\midrule
\textbf{Sexual Models} & 44 (3.4\%)   & 164 (5.3\%) & 208 (4.8\%) \\
\textbf{Deepfake Models}     & 186 (14.5\%) & 1,085 (35\%)  & 1,271 (29\%)  \\
\textbf{Total}            & 1,286 (100\%)  & 3,072 (100\%) & 4,358 (100\%) \\
\bottomrule
\end{tabular}
\caption{Count of Flux models labelled as `Sexual' or `Deepfake' across platforms.}
\label{tab:flux_models}
\end{table*}

\begin{table*}[htbp]
\centering
\begin{tabular}{llrrr}
\toprule
\textbf{Platform} & \textbf{Usage Metric} & \textbf{Sexual} & \textbf{Deepfake} & \textbf{Total} \\
\midrule
\textbf{Hugging Face} & Monthly Downloads & 291,279 (12.6\%) & 75,632 (3.3\%) & 2,311,588 (100\%) \\
& Likes            & 1,187 (8.6\%)  & 221 (1.6\%)  & 13,712 (100\%)  \\
\midrule
\textbf{Civitai}    & Total downloads   & 424,176 (12\%) & 721,705 (21\%) & 3,437,262 (100\%)  \\
& Upvotes          & 32,316 (9\%)    & 59,841 (17\%)   & 344,510 (100\%)  \\
\bottomrule
\end{tabular}
\caption{Flux sexual and deepfake model downloads and likes compared with total on Hugging Face \& Civitai.}
\label{tab:flux_usage}
\end{table*}

Tables \ref{tab:flux_models} and \ref{tab:flux_usage} highlight the extent to which Flux models are fine-tuned for the generation of sexual content and deepfakes. Whilst the total number of explicitly "Sexual" Flux variants is low (4.7\% of total), usage metrics in Table \ref{tab:flux_usage} show that, like Stable Diffusion variants, Flux Sexual Models receive a significant amount of engagement across both Hugging Face and Civitai. Around 12\% of total downloads of Flux variants, or 291,000 downloads a month, are for Sexual Models on Hugging Face. 

Flux additionally appears to be a popular choice for individuals seeking to generate deepfakes. Table \ref{tab:flux_models} shows that 29\% of all assessed Flux variants, or 1,271 models, are designed to generate deepfakes of identifiable individuals. This use case is particularly popular on the Civitai platform, with over a third of all Flux variants analysed intended to generate deepfakes. Table \ref{tab:flux_usage} shows how popular these models are, with deepfake models being downloaded over 700,000 times on Civitai, which is comparable to the total downloads of deepfake Stable Diffusion models on Civitai. 

130 Flux models, predominantly hosted on Civitai, see overlap between Sexual and Deepfake categories. None of these model names use directly sexually explicit terms, but all target the generation of individuals, exclusively women, who feature in adult sexual content, and many hint at adult content by including phrases such as "(18+)" in the title. The remaining deepfake Flux models may be capable of generating NCII, but this use case is not explicitly advertised by the variant creators.

\subsubsection{Stable Diffusion \& Flux perceived gender analysis}
We conduct a perceived gender analysis of the Stable Diffusion and Flux deepfake models analysed to validate the small-scale findings in Part A suggesting that the majority of deepfake models seek to generate images of women. Table \ref{tab:gender_count} confirms this, showing that the overwhelming majority (96\%) of Flux and Stable Diffusion deepfake models across platforms targeted female-presenting individuals. One model sought to generate content for an individual who identifies as non-binary.

\begin{table*}[htbp]
\centering
\begin{tabular}{lrrrr}
\toprule
\textbf{Model} & \textbf{Male} & \textbf{Female} & \textbf{Non-binary} & \textbf{Total} \\
\midrule
\textbf{Stable Diffusion} & 36 (4.4\%)    & 773 (95.4\%)  & 1 (0.1\%)  & 810 (100\%) \\
\textbf{Flux}       & 38 (3.0\%)    & 1,233 (97.0\%) & 0 (0\%)     & 1,271 (100\%) \\
\textbf{Total}      & \textbf{74 (3.55\%)} & \textbf{2,008 (96.4\%)} & \textbf{1 (0.1\%)} & \textbf{2,083 (100\%)} \\
\bottomrule
\end{tabular}
\caption{Count of deepfake models labelled Stable Diffusion and Flux by perceived gender.}
\label{tab:gender_count}
\end{table*}

\begin{table*}[htbp]
\centering
\begin{tabular}{llrrr}
\toprule
\textbf{Platform} & \textbf{Usage Metric} & \textbf{Male} & \textbf{Female} & \textbf{Total} \\
\midrule
\textbf{Hugging Face} & Monthly Downloads  & 1,121 (1.4\%)  & 76,791 (98.6\%) & 77,912 (100\%) \\
& Likes            & 38 (15.8\%) & 203 (84.2\%) & 241 (100\%)   \\
\midrule
\textbf{Civitai} & Total downloads   & 24,364 (1.6\%) & 1,455,532 (98.4\%) & 1,479,896 (100\%) \\
& Upvotes         & 2,353 (2.1\%) & 111,841 (97.9\%)  & 114,194 (100\%) \\
\bottomrule
\end{tabular}
\caption{Usage statistics of deepfake models labelled Stable Diffusion and Flux by perceived gender.}
\label{tab:gender_usage}
\end{table*}

A similar trend can be seen when assessing usage statistics shared by Hugging Face and Civitai in Table \ref{tab:gender_usage}. The overwhelming majority of downloads are for models which target female presenting individuals, with over 1.4 million Flux and Stable Diffusion deepfake variant downloads on Civitai, compared with 24,000 downloads for models targeting male presenting individuals.

\subsubsection{Flux temporal analysis}
\begin{figure*}[htbp]
    \centering
    \includegraphics[width=0.8\textwidth]{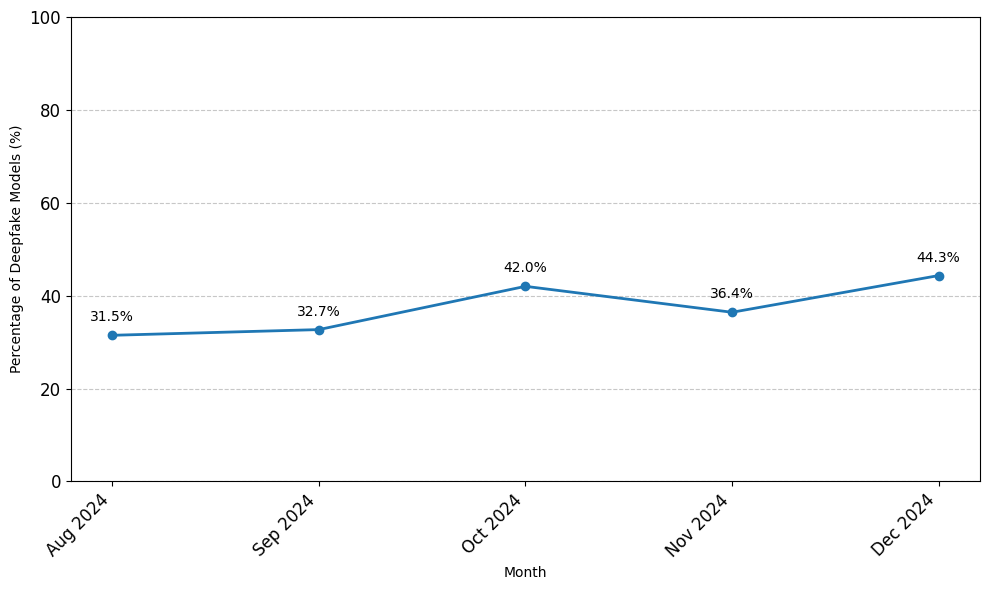}
    \caption{Proportion of Flux LoRA models labelled as `Deepfake' between August 2024 and December 2024 on Civitai platform.}
    \label{fig:fluxtemporal}
\end{figure*}

As Flux models appear particularly popular for deepfake generation since the model’s launch in August 2024 further temporal analysis is conducted to assess the extent to which Flux is used for deepfake generation. Figure \ref{fig:fluxtemporal} shows the proportion of Flux model LoRA variants uploaded to Civitai since August 2024 which were labelled as deepfakes compared to the total number of LoRA Flux models on the platform. As shown in the plot, there has been a steady increase in the proportion of LoRAs targeting identifiable individuals, suggesting that this is a prominent use case for individuals seeking to fine-tune the Flux model, with 44.3\% of all Flux LoRAs in December 2024 seeking to create deepfakes.  

\subsection{Part C Results: Terms of Service and Model Accessibility}
\subsubsection{Terms of Service}
The Terms of Service of Hugging Face and Civitai provides insight into whether these models are normalised on hosting platforms. These Terms of Service give context on whether the platforms discourage deepfakes and sexual content, and whether they require the consent of depicted individuals.

Hugging Face has a published Content Policy which covers rules related to Machine Learning (ML) Artifacts such as models uploaded to the platform \cite{hugging_face_content_2023}. The platform shares a list of prohibited content which covers models uploaded to the Model Hub. Within this policy content is prohibited including:

\begin{enumerate}
    \item[a.] ``\textit{Sexual content used or created for harassment, bullying, or without explicit consent of the people represented}''; and
    \item[b.] ``\textit{Content published without the explicit consent of the people represented}''.
\end{enumerate}

Though many of the models are not explicitly promoted as being designed to generate sexual content, many are published by creators with usernames which suggest such use cases, including terms such as \textit{`playboy'} and \textit{'hotties'}. Similarly, example images provided in model cards of deepfake models contain descriptions including terms such as: `tall, extremely beautiful, fit'. It is therefore likely that many of these models fall foul of policy (a). All of the models depicting identifiable individuals likely violate policy (b), as no references to consent of the individuals depicted were found within any of the model cards examined in this study. Hugging Face do additionally have a reporting mechanism to allow users to request models are taken down, however it appears policy implementation may be reactive, with almost 200 combined Stable Diffusion and Flux deepfake model variants identified on the platform as of December 2024.

Civitai similarly publish a policy pertaining to content released on their platform, and include a section on appropriate depictions of `real people'. Civitai states that \textit{`portraying people in any mature or suggestive context is strictly prohibited'}, and provides examples of prohibitive content relating to clothing, poses and body parts. The policy states that Civitai \textit{`believe that everyone has the right to their own likeness. Unless people have explicitly given consent to be depicted in a certain way, we don't condone it'}. They additionally provide a feedback mechanism for individuals to request removal of their likeness, and say that photorealistic depictions of minors are strictly prohibited. This latter restriction appears well-enforced with no models designed to produce photorealistic depictions of minors identified within this study. Additionally, Civitai have committed to Thorn's Safety by Design principles, which aims to guard against the creation and spread of AI-generated child sexual abuse material \cite{thorn_thorns_2024}. In their latest update Civitai shared that the platform had blocked 400 attempts to upload models optimized for AI-generated Child Sexual Abuse Material, and removed approximately 5-10 models a month. This suggests that takedown mechanisms can be effective at tackling illegal and objectionable content if proactively enforced. 

Similar to the Hugging Face policy, Civitai focuses on the content hosted on the platform itself visible to users, such as example images. However, unlike Hugging Face the policy does not explicitly prohibit models depicting identifiable individuals without consent. This opens a loophole for users seeking to create or generate deepfake sexual content: so long as the images advertising the model on the platform are not sexually explicit, the models do not strictly violate the platform's policy and thus far do not appear to be proactively taken down. This means models uploaded can often infer or suggest sexualised intent indirectly, rather than explicitly acknowledging this purpose via example images, to get around the platform's policy.

The tags and descriptions of deepfake model variants on Civitai provide further support for the existence of this policy gap. Many models are tagged with terms including `nsfw', `sexy', or `porn', suggesting that the model is capable of, or designed for, generating sexually explicit content. A material proportion of models target the generation of individuals who have featured in adult sexual content, or reference popularity on platforms such as OnlyFans, which may suggest the models are intended for adult content. One model description, targeting an adult sexual content actress, describes how the model is `trained by nsfw dataset, so the generated photo [is] mostly bias on NSFW'. Another description describes how `this [model] seems to have a higher chance of revealing clothing/nsfw', acknowledging that models can be used for this purpose. Notably, some models discourage the generation, or publication, of NSFW images, but do not outline any proactive measures taken by the developer to prevent its generation by users such as filters or fine-tuning. As a result, it is likely that many of the models posted on Civitai are intended to, or capable of, generating deepfake sexually explicit content but do not explicitly state this or provide sexually explicit example images, despite such uses being against the intent (but not wording) of the platform's Terms of Service. 

\subsubsection{Model Accessibility}
Thousands of publicly available models spanning a huge range of individuals suggests that the creation and usage of deepfake models has become drastically easier. Insight on this process can be gleaned from the descriptions of models and associated online tutorials. Whilst some of the most popular models cite the use of around 200 images of an individual, many state that they were trained using as few as 20 images of a person. Guidance on Civitai suggests LoRA adapters for the Flux model only require 20 to 30 images, underlining that deepfake models can be created with little new data \cite{redacted_user_detailed_2024}. Scores of tutorials providing guidance on how to conduct LoRA fine-tuning have been released, with one YouTube video advertising that the process can be conducted in as little as 15 minutes and with access to just 24GB of VRAM \cite{comfyui_studio_how_2024, pw_creating_2024, yadav_fine-tuning_2025}. This means that deepfake models can be created of any identifiable individual with a small number of images and time, and with a consumer-grade laptop connected to the internet, posing a risk to public and non-public figures alika.

\section{Discussion}
These results demonstrate that Civitai is frequently used to host and access deepfake models, including model variants targeting Flux and Stable Diffusion models. These deepfake models overwhelmingly target women, some of whom are famous individuals, but who range in popularity. Many models indicate an intention or capability to generate sexually explicit content, suggesting the prevalence of non-consensual intimate imagery generation via these models. Statistics on NCII generation is not possible through these platforms, and equally unlikely to be easily available given the distributed and local nature of model download and usage.

The drastic increase in the number of deepfake models hosted on Hugging Face and Civitai seen in August 2024 suggests that the release of the Flux model during that month marked a seminal moment in making deepfake generators more accessible to the public. The Flux models improved on open model quality, whilst adopting the transformer architecture which enables more customisation of models, via LoRA adapters, at a reduced cost. The ability to fine-tune models efficiently appears to be the primary driver of the rapid rise of deepfake models, with 80\% of `Celebrity' models on Civitai tagged as LoRA adapters. These breakthroughs have lowered the barrier to development, with little time, compute, or expertise now required to create a deepfake model, and with easy paths to share and host these via publicly available online model repositories. 

Despite this concerning situation, steps could be taken to mitigate the risks of further development and proliferation of deepfake models. In the first instance, popular hosting platforms, like Hugging Face and Civitai, could do more to proactively enforce and clarify their Terms of Service. If deepfake models are forbidden on a platform without an individual’s consent, this could be more consistently enforced. This could mean simply deleting models breaking these rules, but these models are often also created by a small group of developers so actions such as banning users could help stem the flow of creation. Platforms could also go further and require models which depict people to explicitly state whether consent was provided by the individuals given and require evidence of consent prior to hosting. The Terms of Service of Civitai could benefit from clarification regarding whether any depiction of individual’s likeness without consent is prohibited on the site, as this is currently unclear. Finally, though non-identifiable sexually explicit content may be permitted, more could be done to identify whether safeguards are in place to prevent the generation of intimate imagery depicting identifiable individuals. This is challenging without direct engagement with the model, but could involve better assessment of metadata or requirements on creators to provide information about fine-tuning data and protections. 

Model creators could also take action to mitigate the risks associated with deepfake imagery. Model developers could invest in mitigation efforts to help identify the provenance of images. Efforts to watermark synthetic media content could help the public and regulators understand which base models are being used for the creation of illegal and objectionable content. Platforms could mandate the provision of watermarks for models uploaded to enable downstream tracking of nefarious uses. However, watermarking may be limited if local users have the ability to remove watermarks once a model is downloaded. Model developers could also seek to invest in further model safety mitigations, for example, by implementing stronger filters on sexual content to reduce the risk of NCII, or greater filtering of sexually explicit content during pre-training. However, the efficacy of these mitigations is limited due to the effectiveness and ease of local fine-tuning.

Regulators could also take a more active role in tackling the dissemination of deepfake models. Currently, regulation focuses primarily on dissemination of images shared without an individual's consent, as well as encouraging transparency of deepfake models. However, if regulatory bodies are empowered to intervene with users who create or use non-consensual models targeting identifiable individuals, this could reduce interest in, and usage of, these models on the mainstream internet. 

\section{Limitations}
These findings do not reflect the complete extent to which deepfakes models are created and disseminated. The results only assess models in the public domain, and many more will likely be created locally and not shared on these platforms. This also means other forms of particularly egregious deepfakes, such as Child Sexual Abuse Content, are not identified within the study. This content is usually consumed and shared in more restricted settings, meaning the phenomenon identified in this paper is likely just a small reflection of the entire picture. As we chose not to generate any images throughout the course of this study in order to avoid contributing to the creation of non-consensual deepfake imagery, our results cannot provide insight into the extent to which the content created by deepfake model variants is sexualised or truly photorealistic. However, many of the models qualitatively reviewed displayed example images which were almost indistinguishable from natural photographs. Our method is also restricted to the limitations of data available via APIs. For the Civitai platform we can measure lifetime model downloads, whereas Hugging Face restricts data only to monthly downloads, which does not allow a direct comparison of the platforms, potentially casting Hugging Face in a more positive light than Civitai. Our methodology is also reliant on accurate user tagging of models on Civitai as `Celebrity' models, which could lead to false positives or negatives. Finally, reviewed models may be duplicated on both Hugging Face and Civitai, though we chose to include these potential duplicates within the detailed analysis to assess the popularity of deepfakes independently on both platforms. 

\section{Conclusion}
This paper has provided empirical evidence charting the rise of non-consensual deepfake image generators, and their use in generating sexual content. This paradigm shift has been driven by improved accessibility and a proliferation of deepfake image generators, aided by the model sharing sites such as Civitai and high quality, easily fine tuneable, open access base models like Flux. 

We identify almost 35,000 deepfake models, predominantly hosted on the Civitai platform and garnering just under 15 million downloads cumulatively since the platform’s launch in November 2022. Of over 2,000 deepfake Flux and Stable Diffusion models analysed in detail we found 96\% targeted women. We find deepfake models target individuals ranging from celebrities to Instagram users without significant followings. Additional metadata analysis provides evidence that many deepfake models intend to enable direct generation of sexual content, or non-consensual intimate imagery. While model hosting platforms have Terms of Service that ostensibly prohibit such content, our findings suggest that loopholes exist and enforcement is lacking, with deepfake models capable of generating NCII available on demand. Our results are likely a fraction of the full story. Each model variant downloaded can be used to create an unlimited number of deepfake images, and many malicious actors may not upload models to public repositories, instead storing models locally or sharing via private channels. The extent of this issue points to the need for greater public discussion and warrants intervention from model creators, hosting sites, and regulators to address the rise of AI-generated NCII. 

\begin{acks}
The authors would like to thank the following individuals for helpful discussions and feedback throughout the course of this project: Inga Campos, Sasha Brown, John Buckley, Susanna Ricco, Susan Hao, Julia Haas, Will Babbington, Kevin McKee, and Scott Pope.

This study was reviewed and received approval from the University of Oxford’s Central University Research Ethics Committee, application ID 987316. Brent Mittelstadt and Chris Russell’s contributions to this work have been supported through research
funding provided by the Wellcome Trust (grant nr 223765/Z/21/Z), Sloan Foundation (grant nr G2021-16779), Department of Health and Social Care, EPSRC (grant nr EP/Y019393/1), and Luminate
Group. Their funding supports the Trustworthiness Auditing for AI project and Governance of
Emerging Technologies research programme at the Oxford Internet Institute, University of Oxford.
During the course of this work Will Hawkins held an employed position at Google DeepMind.
\end{acks}

\bibliographystyle{ACM-Reference-Format}
\bibliography{t2i-bib2}

\appendix

\section{Red flag analysis details}
Table 11 outlines the types of red flags identified which suggest deepfake model variants (tagged by creators as "Celebrity" models or manually identified as a deepfake model during this study) are capable of generating sexually explicit content.

\label{appendix:d}

\begin{table*}[htbp]
\centering
\begin{tabular}{|p{0.25\textwidth}|p{0.7\textwidth}|}
\hline
\textbf{Red Flag Type} & \textbf{Detail} \\
\hline
Tags & Tags directly suggesting intimate imagery is possible using the model. E.g. ``sex'', ``sexy''. For example see one \href{https://civitai.com/models/17873/emma-watson}{Emma Watson LoRA}. The tag ``sex'' or ``sexy'' appears over 5,000 times across the 34,440 Celebrity models.

Other models which indicate pornographic content of identifiable people includes: ``porn'' (221 tags) and ``breasts'' (85 tags). \\
\hline
Usernames & Usernames which suggest content created are targeting sexualised content. E.g. one prominent Hugging Face uploader featured the term ``Playboy'' in their name. \\
\hline
Targeted individuals featuring in pornography & Many models explicitly target individuals who have featured in pornographic content. 2,975 of the 34,440 Celebrity models featured the tag ``pornstar'', suggesting the variant is intended for the generation of sexual content. Other models feature "pornstar" within the title, or indicators such as the term "(18+)". \href{https://civitai.com/models/58815?modelVersionId=1099895}{Example variant}. \\
\hline
Descriptions & Model descriptions often discuss use cases, such as pornographic generation. The term ``porn'' is referenced 765 times across the 34,440 Civitai Celebrity tagged models, ``sex'' 722 times, and ``breast'' 446 times. See example \href{https://civitai.com/models/441491?modelVersionId=1158331}{Electra 90s LoRA} description.

Some of these references expressly discourage pornographic generation, but do not appear to attempt any technical mitigations to prevent this \href{https://civitai.com/models/22787?modelVersionId=27212}{(example)}. \\
\hline
\end{tabular}
\caption{Red flags identified which indicate deepfake models are capable of generating sexually explicit content}
\label{tab:red_flags}
\end{table*}

\section{Model analysis filtering criteria}
\label{appendix:a}
When identifying Stable Diffusion and Flux models we searched the Hugging Face and Civitai APIs. For Stable Diffusion models we searched using the terms `Stable Diffusion' or `SD' in the model name. For Flux models we searched using the term `Flux' in the model name.

On Hugging Face, we filtered out models with fewer than 10 monthly downloads, to remove inactive models. On Civitai, where lifetime, rather than monthly, downloads are tracked, we filtered out models with fewer than 250 downloads.

Across both platforms we manually removed models which were uploaded by the model developer (i.e. Stability AI and Black Forest Labs), and these represent base models, rather than new variants. We remove quantisations of models for the same reason. 

As we identify during the platform analysis that 80\% of deepfake models are created using LoRA, and the Civitai platform allows filtering for this model type, on Civitai we analyse only models labelled as `LoRA' by model creators. 

Based on model name comparisons we identify 716 potential duplicates hosted on both the Hugging Face and Civitai platform. These potential duplicates are retained within the overall dataset in order to analyse popularity between platforms.

\section{Model analysis labelling criteria}
\label{appendix:c}
For the manually labelled Stable Diffusion and Flux models the following terms were searched in the titles of models to determine whether a model variant was intended for sexually explicit use ('Sexual Model'): `nsfw', `porn', `sexy', `babes', `hentai', `unsafe', `xxx'. Additionally, we included any deepfake models which targeted individuals featuring in adult sexual content within the sexually explicit category.

To determine whether a reviewed Stable Diffusion or Flux model was a 'deepfake model' we assessed model titles for indicators that an identifiable person was intended to be generated. If names were included in the titles, we conducted Google Searches to determine if the individual was identifiable. We included models where an individuals name is not mentioned but their identity is inferred, such as a TV or Film character ("Indiana Jones"). For these models we also reviewed whether the example images were seeking photorealistic content, and removed models which only sought cartoon depictions of characters. 

\section{Code}
\label{appendix:b}
The code used to access the Civitai and Hugging Face APIs and download model can be found at the following GitHub repository: https://github.com/WillHawkins3/deepfakesondemand 

\end{document}